\documentclass[aps,prl,preprint,groupedaddress,floatfix]{revtex4-1}
\usepackage{graphicx}
\usepackage{bm}

\begin{document}

%\preprint{}

\title{On the 3-D structure and dissipation of reconnection-driven flow-bursts}


\author{J. F. Drake}
%\thanks{}
\affiliation{University of Maryland, College Park, MD 20742, USA}
%\altaffilitiation{University of California, Berkeley, CA 94720, USA}
\author{M. Swisdak}
\affiliation{University of Maryland, College Park, MD 20742, USA}
\author{P. A. Cassak}
\affiliation{West Virginia University, Morgantown, WV 26506, USA}
\author{T. D. Phan}
\affiliation{University of California, Berkeley, CA 94720, USA}


\date{\today}

\begin{abstract}
The structure of magnetic reconnection-driven outflows and their
dissipation are explored with large-scale, 3-D particle-in-cell (PIC)
simulations. Outflow jets resulting from 3-D reconnection
with a finite length x-line form fronts as they propagate into the
downstream medium. A large pressure increase ahead of this
``reconnection jet front'' (RJF), due to reflected and transmitted
ions, slows the front so that its velocity is well below the velocity
of the ambient ions in the core of the jet. As a result, the RJF slows
and diverts the high-speed flow into the direction perpendicular to
the reconnection plane. The consequence is that the RJF acts as a
thermalization site for the ion bulk flow and contributes
significantly to the dissipation of magnetic energy during
reconnection even though the outflow jet is subsonic. This behavior
has no counterpart in 2-D reconnection. A simple analytic model
predicts the front velocity and the fraction of the ion bulk flow
energy that is dissipated.
\end{abstract}

%\keywords{}

\maketitle

%\section{Introduction\label{intro}} 

Magnetic reconnection is the dominant mechanism for dissipating
magnetic energy in large-scale plasma systems and is the driver of
explosive events such as flares in astrophysical systems and flow
bursts in the Earth's magnetosphere. Most of the energy released
during reconnection takes place not at the x-line but downstream in
the exhaust where newly reconnected field lines relax their magnetic
tension. In the MHD description the Petschek shocks that bound the
exhaust both drive the Alfv\'enic exhaust and heat the upstream plasma
entering the exhaust \citep{Petschek64,Lin95}. In the nearly
collisionless environment of many systems, while the exhaust outflow
is close to the MHD prediction \citep{Sonnerup81}, the exhaust heating
results from counterstreaming ions \citep{Hoshino98,Gosling05} rather
than Petschek shocks. The kinetic energy of the bulk flow driven
during reconnection is a substantial fraction of the released magnetic
energy and in natural systems this energy is ultimately
dissipated. However, the dominant processes that control the
dissipation of these flows and their universality have not yet been
established.

During solar flares the termination shock that has been observed at
the low altitude edge of coronal reconnection exhausts
\citep{Masuda94} is a possible mechanism for the dissipation of the
energy in the bulk flow. Supra Arcade Downflows (SADs)
\citep{McKenzie99}, which are believed to be driven by reconnection,
are observed to slow during their downward trajectory toward the solar
surface. In the Earth's magnetotail a key observational discovery was the
formation of narrow boundary layers or fronts at the interface of the
high-speed reconnection jets and the essentially stationary ambient
plasma downstream. At the front the
amplitude of the magnetic field normal to the initial current layer
($B_z$ in the magnetotail) increases abruptly
\citep{Ohtani04,Runov09,Runov11}. Initially such fronts (dubbed
``dipolarization fronts'') were believed to result from the slowing of
the reconnection outflow as it impacted the strong dipole field of the
Earth, similar to coronal termination shocks. However, the measured
propagation of these fronts over large distances both Earthward and
tailward of the reconnection site \citep{Ohtani04,Angelopoulos13} is
strong evidence that these fronts are generically associated with the
development of reconnection in natural systems and are not specific to
the geometry of a particular system.  

The role that reconnection jet fronts (RJFs) play is, however,
unclear. It has been suggested that RJFs may be important sites for
energy dissipation \citep{Hoshino01b,Runov11a,Angelopoulos13}. A
number of 2-D reconnection simulations (corresponding to an infinite
length x-line) have been carried out to explore the structure of RJFs
\citep{Sitnov09,Wu12}. On the other hand, it is unlikely that
reconnection in physical systems is 2-D since reconnection very likely
onsets in a spatially localized region.  Flow-bursts and associated
RJFs in the magnetotail are localized in the cross-tail ($y$)
direction with characteristic scales of several Earth radii $R_E$
\citep{Angelopoulos97,Nakamura04} and therefore correspond to finite
length x-lines. SADs have similarly been interpreted as resulting
from reconnection with finite length x-lines
\citep{Linton06,Cassak13}.  We show that the structure of the exhaust
and its dissipation depends critically on its 3-D structure.

Since we are focusing on the structure of the reconnection outflows
and the associated RJF and not on the structure of the dissipation
region where field lines change topology, we explore the dynamics with
the Riemann approach \citep{Lin95}. Consistent with the observations,
we study how reconnection develops in a 3-D model with a finite x-line
by imposing a spatially localized region of reconnected flux $B_z$
on top of a Harris current sheet. A PIC model is
used so that the collisionless dissipation of reconnection-driven
flows can be studied. The system is initially in pressure balance but
the curvature forces that drive reconnection are unbalanced and drive
the outflow. A large pressure increase ahead of the front, due to
reflected and transmitted ions, slows the front so that its velocity
is well below the velocity of the ambient reconnection outflow. As a
result, the front slows and diverts the outflow into the ion drift
direction of the downstream current layer. The front, therefore acts
as a thermalization site for the energy of the flow. A simple
model illustrates how the energy in the flow is dissipated and is
proposed as a prototype for understanding how reconnection-driven
flows are dissipated in nature. No significant enhancement in electron
dissipation takes place at the front.

\begin{figure*}
 \includegraphics[{width=8.5cm}]{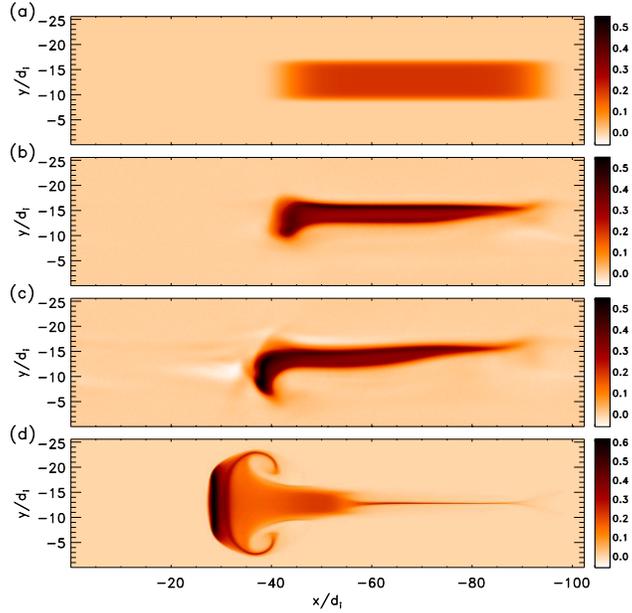}
 \caption{\label{by} (Color online) From the PIC simulation plots of
   $B_z$ in the $x-y$ plane at the center of the current sheet ($z=0$) at (a) $t=0$, (b) $t=12\Omega_{ci}^{-1}$ and (c)
   $t=24\Omega_{ci}^{-1}$. In (d) a similar plot from an MHD
   simulation with nearly identical initial conditions at the same
   time as in (c).}
 \end{figure*}

We explore a system periodic in three directions: with $x-z$ the
plane of reconnection and the reconnection outflow along
$x$. Superimposed on a double Harris current layer $B_x(z)$ with a
half-width of $2.0d_i$ (with $d_i$ the ion inertial length) is a
region of uniform magnetic flux $B_z(x,y)$ that is localized in the
$x-y$ plane as shown in Fig.~\ref{by}(a). The density in the region of
$B_z\neq 0$ is reduced to the background density $n_0$ of the Harris
system. The electron and ion temperatures are adjusted so that the
total pressure is balanced with $T_i/T_e=5$, which is typical for the
magnetosphere. Required currents are carried by both species, in
proportionality to their temperature. Unbalanced forces associated
with magnetic tension will drive the plasma in the region of $B_z\neq
0$ to the left in Fig.~\ref{by}(a). The results of our PIC simulations
are presented in normalized units: the magnetic field to the
asymptotic value $B_0$ of the Harris reversed field, the density to
the value at the center of the current sheet minus $n_0=0.3$,
velocities to the Alfv\'en speed $c_A$, lengths to $d_i$, times to the
inverse ion cyclotron frequency $\Omega_{ci}^{-1}$, and temperatures
to $m_i c_A^2$. The computational domain is $102.4d_i\times
25.6d_i\times 25.6d_i$. Other parameters of the simulations are a
mass-ratio $m_i/m_e=25$, which is sufficient to separate the dynamics
of the two species \citep{Hesse99}, and speed of light $c=15c_A$.

\begin{figure*}
 \includegraphics[{width=8.5cm}]{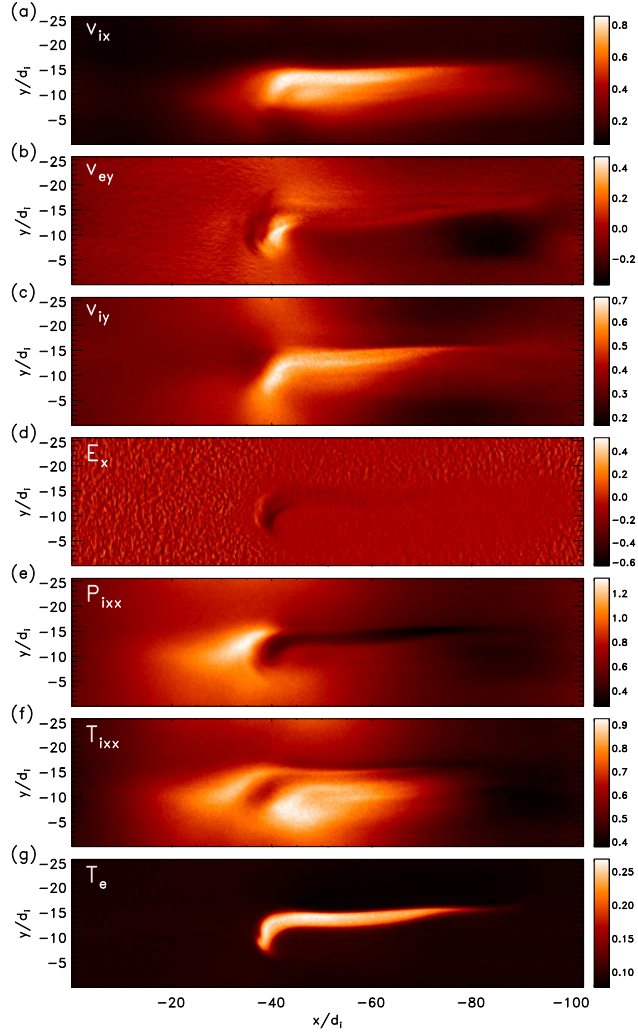}
 \caption{\label{2dplots} (Color online) As in Fig.~\ref{by} at $t=24\Omega_{ci}^{-1}$ plots of (a)
   $v_{ix}$, (b) $v_{ey}$, (c) $v_{iy}$, (d) $E_x$, (e) $p_{ixx}$, (f)
   $T_{ixx}$ and (g) $T_e$.}
 \end{figure*}

Shown in Fig.~\ref{by} is $B_z$ in the center of the current sheet ($z=0$) at
(a) $t=0$, (b) $t=12\Omega_{ci}^{-1}$ and (c)
$t=24\Omega_{ci}^{-1}$. The outflow carries the flux $B_z$ to the left
propelled by the unbalanced magnetic curvature forces that drive
reconnection.  The corresponding flow $v_{ix}$ at
$t=24\Omega_{ci}^{-1}$ is shown in Fig.~\ref{2dplots}(a). A surprise
in this data is the strong positive flow just below the main flow in a
region where $B_z\sim 0$ and there is therefore no curvature. The
reason for this flow is discussed later. In the core of the flow jet
the flux $B_z$ is carried upward in Fig.~\ref{by}, which is in the
electron drift direction (Fig.~\ref{2dplots}(b)), and is compressed at
the upper edge.  In contrast, the left edge of the jet turns towards
downward, which is in the ion drift direction
(Fig.~\ref{2dplots}(c)). Similar turning in the ion drift direction
was seen in 3-D PIC simulations of interchange turbulence in the
magnetotail \citep{Pritchett13}. The motion downward at the front of
the jet is due to an electric field $E_x$ directed to the right which
causes both species to drift downward
(Figs.~\ref{2dplots}(b)-(d)). The asymmetry of the structure of the
jet in the y-direction is contrasted with the results of an MHD simulation
with nearly identical initial conditions (the minimum
density is slightly higher in the MHD case) in which the jet forms a
symmetrical structure in the $x-y$ plane (Fig.~\ref{by}(d)). These MHD
flows are similar to those from earlier 3-D MHD simulations of plasma
interchange-driven flows in the magnetotail \citep{Birn04b}.

\begin{figure*}
 \includegraphics[{width=8.5cm}]{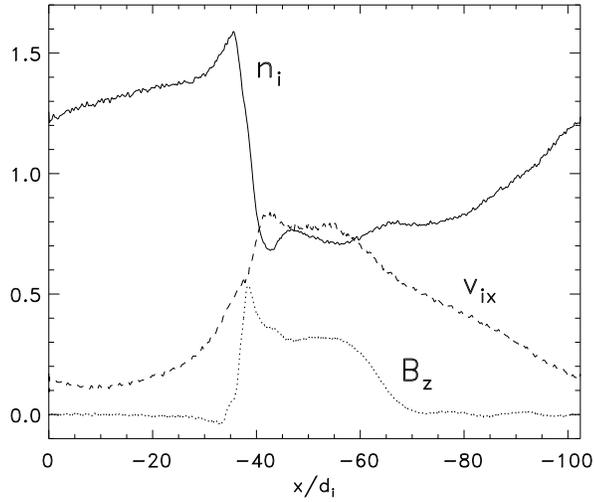}
 \caption{\label{cuts} Cuts of the ion density $n_i$ (solid), $v_{ix}$ (dashed) and $B_z$ (dotted) along $x$ at $z=0$ and $y=-11.3d_i$ at the same time as in Fig.~\ref{2dplots}.}
 \end{figure*}

In Fig.~\ref{cuts} are cuts of the ion density (solid), $B_z$ (dotted)
and the ion velocity $v_{ix}$ (dashed) versus $x$ in the center of the
current sheet at $t=24\Omega_{ci}^{-1}$. The density drops sharply
across the front and into the jet although the density minimum of
around $0.7$ is well above the initial condition of $0.3$. $B_z$ rises
sharply across the front and exhibits the distinctive dip and
overshoot that are often seen in the observations
\citep{Ohtani04,Runov09}. The overshoot results from local compression
at the front, which produces a similar peak in the density. The region
of negative $B_z$ ahead of the front can also be seen as the white
region in Fig.~\ref{by}(c) in the interval $-35d_i < x < -25d_i$ and
$-9d_i>y>-13d_i$. The mechanism for this reversal in $B_z$ appears to
be similar to the self-generation of magnetic fields in the Weibel
instability and will be discussed more fully in a separate
publication.

%Briefly, in the
%frame of the front ions and electrons to the left of the front in
%Fig.~\ref{by}(c) are moving together toward the front. A negative
%perturbation of $B_z$ ahead of the front deflects the electrons
%downward while the ions because of their higher mass continue toward
%the front. The downward motion of electrons produces an upward current
%that combines with the downward current from ions closer to the front
%to form a clockwise current loop that re-inforces the region of
%$B_z<0$ ahead of the front in Fig.~\ref{by}(c).

\begin{figure*}
 \includegraphics[{width=8.5cm}]{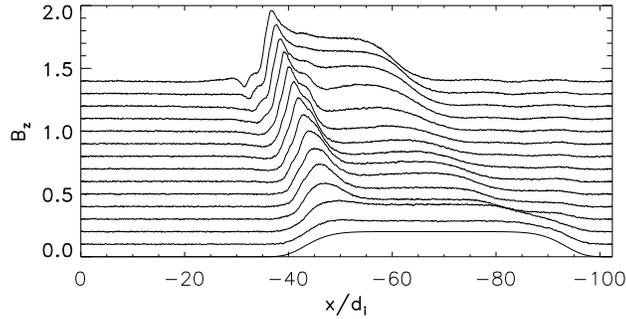}
 \caption{\label{stack} A stack of cuts of $B_z$ along $x$ at $z=0$ and $y=-11.3d_i$ separated by time intervals of
   $2\Omega_{ci}^{-1}$ starting from $t=0$. The velocity of the front
   calculated from the peaks of $B_z$ is $0.46c_A$.}
 \end{figure*}

\begin{figure*}
 \includegraphics[{width=8.5cm}]{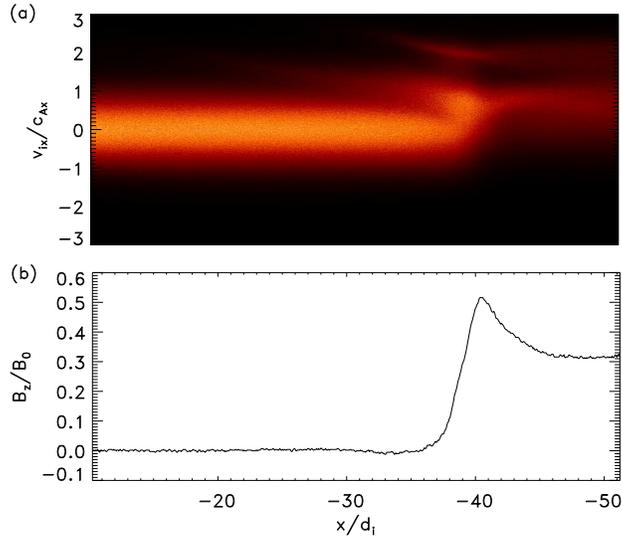}
 \caption{\label{phase} (Color online) In (a) the ion phase space $v_x-x$ in the
   center of the current sheet at $y=-11.3d_i$ and $t=20\Omega_{ci}^{-1}$. In (b) a cut of $B_z$ at the same time and location.}
 \end{figure*}

The ion velocity rises gradually ahead of the front as in observations
and reaches a plateau around $0.8c_A$, which is well below that
expected based on the upstream Alfv\'en speed ($1.8c_A$). The velocity
$v_f$ of the front is around $0.46c_A$ and is calculated by stacking
cuts of $B_z$ versus $x$ at several times (Fig.~\ref{stack}). The
reduced velocity of the front compared with that of the core of the
jet results from the buildup of ion pressure ahead of the front
shown in Fig.~\ref{2dplots}(e). The increase in pressure is largest at
the top corner of the front and serves to both slow the jet and
deflect it downward.  Such pressure enhancements have been documented
in satellite measurements \citep{liu13a}. The pressure increase is a
consequence of the reflection of ions in the current sheet off of the
head of the front, which has also been documented in satellite
observations \citep{Zhou10} and discussed in 2-D reconnection models
\citep{Wu12}. The penetration of high velocity ions in the jet
through the front also contributes to the pressure increase. Both
classes of particles can be seen to the left of the front in the $x-v_x$
phase space in Fig.~\ref{phase}(a), which is from the center of the
current sheet with $y=-11.8d_i$ and $t=20\Omega_{ci}^{-1}$. The cut of
$B_z$ in Fig.~\ref{phase}(b) shows the location of the front. As
expected, the reflected ions have a velocity close to $c_A$, which
is around twice $v_f$. In Fig.~\ref{2dplots}(f) is the ion temperature
$T_{ixx}$ corresponding to the pressure in Fig.~\ref{2dplots}(e).  The
increase in ion temperature associated with the reflected ions is
evident.

The mechanism that produces the enhanced ion temperature below the
jet and to the right of the front is also responsible for the
strong leftward-directed flow in the same region shown earlier in
Fig.~\ref{2dplots}(a).  Ions moving to the left in the core of the jet also
drift downward (Fig.~\ref{2dplots}(c)) and eventually
exit the region of strong $B_z$ into the adjacent stationary
plasma. These ions continue to flow to the left through the stationary background ions. The resulting
counterstreaming ion velocity distributions in this region where
$B_z\sim 0$ have a net drift to the left (Fig.~\ref{2dplots}(a)) and
produce an effective $T_{ixx}$ as seen in Fig.~\ref{2dplots}(f). A cut
along $x$ at $y=-10d_i$ (not shown) reveals a localized peak in $B_z$
at the front but with a flow to the left that remains large well to the
right of the front where $B_z\sim 0$. Such behavior has been
documented in THEMIS magnetotail observations \citep{Runov11a}. In
contrast with the ions, we have measured no significant increase in
the electron temperature at the front (Fig.~\ref{2dplots}(g)) in spite
of the intense electron current in the $x-y$ plane that produces the
rather complex magnetic structure $B_z$ in Fig.~\ref{by}.

Before further addressing the properties of the front, we emphasize
that this sharp boundary is not a shock. First, in the upstream region
to the right of the front the fast-mode phase speed based on the total
plasma and magnetic pressure is around $1.0c_A$ while the flow speed
of the jet is around $0.8c_A$ so the upstream Mach number is less than
unity. Further, if the front were a shock, the flow through the shock
would carry the flux $B_z$ across the shock into the downstream
region, which is not seen in the simulation data.  Nevertheless, while
the front is not a shock, that the velocity in the core of the
flow-burst is substantially higher than that of the front has
important implications for understanding the dissipation of
reconnection driven flows. The plasma within the jet eventually
catches up to the front where it is compressed and deflected with
reduced velocity downward in Fig.~\ref{2dplots}(c). The consequence is
that the integrated volume of plasma behind the flow-burst (measured
by the reduction in the size of the non-zero $B_z$ region) decreases
with time. This can be seen in Figs.~\ref{by}(b)-(c). The integrated
magnetic flux $B_z$ in the jet is decreasing with time and the
corresponding flux convecting downward is increasing (Fig.~\ref{by}). Thus, the
front is more than simply the front edge of the jet. Rather, it
is the site for conversion of flow energy into ion thermal energy and
much of the plasma that makes up the jet will be directly
processed within the front.

A simple analytic calculation illustrates how the reduction in flow
energy takes place. We consider a simple 2-D system in
the $x-y$ plane in which the plasma in the current sheet (density
$n_{cs}$) interacts with the plasma in the jet (density $n_b$).
The front and jet velocities are $v_f$ and $v_b$, respectively,
while the current sheet ions are at rest. For simplicity, we ignore
the ambient drift along $y$, which adds complexity to the calculation but
does not change the final result. In the frame of the front, ions in
the current sheet move with a velocity $-v_f$ along $x$, are reflected
by the magnetic boundary and leave the front with a velocity
$v_f$. The jet ions have an incident velocity $v_b-v_f$ and are
deflected into the $y$ direction with their speed unchanged.  Force
balance at the front, neglecting the residual magnetic stress, yields
$2n_{cs}v_f^2=n_b(v_b-v_f)^2$, which can be solved for the front
velocity $v_f=Rv_b/(1+R)$ where $R=\sqrt{n_b/2n_{cs}}$. In the frame
of the front, neither the current sheet nor the jet ions change
energy. In the simulation frame, however, there is a transfer of energy
from the jet ions to the current sheet ions. The
change in energy $\Delta W=W_f-W_i$ of the jet ions is
given by
\begin{equation}
\Delta W=\frac{1}{2}m_in_b(v_f^2+(v_b-v_f)^2-v_b^2)=-2W_i\frac{R}{(1+R)^2}.
\label{deltaw}
\end{equation}
with $W_i=m_in_bv_b^2/2$. This energy loss corresponds to the energy
gain of the current sheet ions. Thus, the fraction of energy
conversion is linked to the density ratio between the jet and current
sheet ions. In the limit of $n_b/n_{cs}\rightarrow 0$ there is no
energy conversion. For a typical value $n_b/n_{cs}=0.2$
\citep{Runov09} the fraction of energy conversion is $0.37$ according
to this simple model. The model, of course, greatly simplifies a very
complex system. The model prediction of the front velocity in the
simulation is $0.3c_A$ compared with the measured value of
$0.46c_A$. Finally, we emphasize that the deflection of the jet into
the $y$ direction has no counterpart in a 2-D reconnection model.

Estimates of the front velocity based on multi-spacecraft THEMIS
observations have been presented \citep{Runov11} but a direct
comparison with the velocity of the core of the jet has not
been carried out. This comparison would be facilitated if a local
measure of the velocity of the front from single spacecraft data could
be obtained. From the simulation we have evaluated the local ${\bf
  E}\times {\bf B}$ velocity at the peak of $B_z$. At
$t=30\Omega_{ci}^{-1}$ this velocity is $0.47c_A$, which is quite
close to the value of $0.46c_A$ deduced from the stack of plots in
Fig.~\ref{stack}. The local ion velocity $v_{ix}$ at the peak of $B_z$
is also close to but somewhat higher than the front velocity. 

The overshoot in $B_z$ seen in many observations of RJF encounters in
the magnetotail is already evidence of the pileup of the jet
plasma at the front but direct comparisons of the front velocity with
that of the core of the jet are necessary to test the ideas
presented here. While the focus of the present paper is on
reconnection-driven jets, jets driven by the magnetized
Rayleigh-Taylor instability \citep{Birn04} or other mechanisms might
also exhibit similar deflections and associated dissipation of bulk
flow energy.

%\subsection{}
%\subsubsection{}

% If in two-column mode, this environment will change to single-column
% format so that long equations can be displayed. Use
% sparingly. 
%\begin{widetext}
%\end{widetext}

% figures should be put into the text as floats.
% Use the graphics or graphicx packages (distributed with LaTeX2e)
% and the \includegraphics macro defined in those packages.
% See the LaTeX Graphics Companion by Michel Goosens, Sebastian Rahtz,
% entire page. There is no need to do explicit centering.

\begin{acknowledgments} The authors would like to thank V.\ Angelopoulos and A.\ Runov and L.\ S.\ Shepherd for helpful discussions. This work was
supported in part by NASA grants NNX14AC78G, NNX08AO83G and NNX10AN08A
and NSF Grants ATM-0903964 and AGS-093463. Computations were performed
at the National Energy Research Scientific Computing Center and on
NASA Advanced Supercomputing Division resources.

\end{acknowledgments}

% Surround figure environment with turnpage environment for landscape
% figure
% \begin{figure}
% \end{figure}

% tables should appear as floats within the text
% The ruledtabular enviroment adds doubled rules to table and sets a
% reasonable default table settings.
% Use the table* environment to get a full-width table in two-column
% Add \usepackage{longtable} and the longtable (or longtable*}
% environment for nicely formatted long tables. Or use the the [H]
% placement option to break a long table (with less control than 
% \begin{table}%[H] add [H] placement to break table across pages
% \begin{ruledtabular}
% Lines of table here ending with \\
% \end{tabular}
% \end{ruledtabular}
% \end{table}

% Surround table environment with turnpage environment for landscape
% table
% \begin{table}
% \begin{ruledtabular}
% \end{tabular}
% \end{ruledtabular}
% \end{table}

%\appendix
%\section{}

\newpage
%\bibliographystyle{apsrev}
%\bibliography{bib}

\end{document}